\documentclass[runningheads]{llncs}
\usepackage{graphicx}
\usepackage{color,soul}
\usepackage{array}
\newcolumntype{P}[1]{>{\centering\arraybackslash}p{#1}}

\begin{document}
\title{Sickle Cell Disease Severity Prediction from Percoll Gradient Images using Graph Convolutional Networks}

\author{Ario Sadafi\inst{1,2} \and
Asya Makhro\inst{3} \and
Leonid Livshits\inst{3} \and
Nassir Navab \inst{2,4} \and \\
Anna Bogdanova\inst{3} \and
Shadi Albarqouni \inst{2,5} \and
Carsten Marr \inst{1}}
%(carsten.marr@helmholtz-muenchen.de)
% index{Sadafi, Ario}
% index{Makhro, Asya}
% index{Livshits, Leonid}
% index{Navab, Nassir}
% index{Bogdanova, Anna}
% index{Albarqouni, Shadi}
% index{Marr, Carsten}
\institute{Institute of Computational Biology, Helmholtz Zentrum München - German Research Center
for Environmental Health, Neuherberg, Germany \and
Computer Aided Medical Procedures, Technical University of Munich, Germany
\and
Red Blood Cell Research Group, Institute of Veterinary Physiology, Vetsuisse Faculty and the Zurich Center for Integrative Human Physiology, University of Zurich, Zurich, Switzerland
\and
Computer Aided Medical Procedures, Johns Hopkins University, USA
\and
Helmholtz AI, Helmholtz Center Munich, Neuherberg, Germany}
\authorrunning{A. Sadafi et al.}
\titlerunning{SCD Severity Prediction from Percoll Gradient Images}
\maketitle              % typeset the header of the contribution
\begin{abstract}
Sickle cell disease (SCD) is a severe genetic hemoglobin disorder that results in premature destruction of red blood cells. Assessment of the severity of the disease is a challenging task in clinical routine, since the causes of broad variance in SCD manifestation despite the common genetic cause remain unclear. Identification of the  biomarkers that would predict the severity grade is of importance for prognosis and assessment of responsiveness of patients to therapy. Detection of the changes in red blood cell (RBC) density by means of separation of Percoll density gradient could be such marker as it allows to resolve intercellular differences and follow the most damaged dense cells prone to destruction and vaso-occlusion. Quantification of the images obtained from the distribution of RBCs in Percoll gradient and interpretation of the obtained is an important prerequisite for establishment of this approach.
Here, we propose a novel approach combining a graph convolutional network, a convolutional neural network, fast Fourier transform, and recursive feature elimination to predict the severity of SCD directly from a Percoll image. Two important but expensive laboratory blood test parameters measurements are used for training the graph convolutional network. To make the model independent from such tests during prediction, these two parameters are estimated by a neural network from the Percoll image directly. On a cohort of 216 subjects, we achieve a prediction performance that is only slightly below an approach where the groundtruth laboratory measurements are used.
Our proposed method is the first computational approach for the difficult task of SCD severity prediction. The two-step approach relies solely on inexpensive and simple blood analysis tools and can have a significant impact on the patients' survival in underdeveloped countries where access to medical instruments and doctors is limited. 

\end{abstract}
\section{Introduction}
Sickle cell disease (SCD) is a disorder caused by mutations in position 6 of $\beta$-hemoglobin gene (hemoglobin S) with extreme variability at phenotypic level. In some patients the disease manifestation is so mild that they remain asymptomatic most of the time while others die before the age of five from several of the severe complications associated to the SCD \cite{sebastiani2007network}. 
Individuals who have the hemoglobin S variant are naturally protected against malaria, which has a profound influence on the spread of sickle cell disease globally affecting the tropical (African and Asian) countries the most. Many of these countries are not able to support diagnosis and appropriate healthcare for this group of patients leading to a drop in the life expectancy from 45-55 years in high income countries to 90\% death rate before the age of 5 in low income countries \cite{wastnedge2018global}.

Severity monitoring and prediction of the SCD is therefore an important task along with development of new effective and inexpensive therapeutic strategies. Changes in severity allow monitoring of the treatment efficiently and for prediction and prevention of life-threatening complications in short future.
To date, there is no practical test based on red blood cells (RBCs) density separation analysis available for prediction of the severity of the disease for a patient.

An important parameter for disease severity assessment is the percentage of hypo- and hyperchromic cells. RBCs with a hemoglobin concentration above 410 g/l are called hyperchromic and characterized by low cellular deformability \cite{deuel2012asymptomatic} and increased probability of aggregation of the hemoglobin S which directly associates with advanced severity and poor prognosis for the SCD patients \cite{de2009pathophisiology,brugnara2013red}. In contrary, Hypochromic RBCs with low hemoglobin content are associated with a lower probability of hemoglobin S aggregation and sickling and thus with mild disease manifestation. 
Measurement of these parameters using blood smears is laborious and time-consuming, when done manually by skilled personnel, or rather requires expensive medical laboratory equipment, when automated.

The spleen plays an important role in clearing the blood from old, broken, dehydrated or hyperchromic red blood cells (RBCs). A normal and functioning spleen reduces the intravascular hemolysis of damaged cells (where cells rupture in the blood vessels) and prevents vaso-occlusive crisis (where terminally dense sickle cells block circulation of blood vessel leading to painful crisis) and vascular damage in SCD patients \cite{brousse2014spleen}. However, fibrosis and progressive atrophy of the spleen resulting finally in necrosis of the organ, known as autosplenectomy, which is often observed in SCD patients with severe disease phenotype. It is a known problem in children with SCD due to repeated splenic vaso-occlusive events in the organ \cite{brousse2014spleen}. Measuring spleen size with ultrasound is a common way to evaluate the organ's condition in SCD. 

We here propose a computational approach that circumvents expensive lab tests and relies solely on the measurement of spleen size and a Percoll image.  Percoll images are used to assess the density of the cells and particles. After centrifugation, several bands with different thicknesses are formed by RBCs of similar density (see Fig. \ref{overview}) holding important information about a SCD patient's condition. Back in 1984, Fabry et al. \cite{fabry1984objective} observed a decrease in the dense fraction of Percoll images in SCD patients suffering from painful crisis in 11 patients over 14 painful crisis image. This information can also be computationally analyzed:
Sadafi et al. \cite{isbiPercoll} introduced a hybrid approach based on CNNs and features extracted from fast Fourier transform to classify a variety of hereditary hemolytic anemias using Percoll image data.

To predict the severity of the SCD patients, we are proposing an approach based on graph convolutional networks (GCN) to form a population graph \cite{parisot2017spectral} on our data. The similarity of the GCN edges is calculated using lab (Percentage of hypo- and hyperchromic RBCs) and clinical data (spleen size). 
The spleen size is measured using ultrasound. We propose a CNN based approach to have an easy to access and affordable anywhere in the world way to estimate required lab data from the Percoll image.

\section{Methodology}
\begin{figure}[t]
    \centering
    \includegraphics[width=\textwidth, page=1, trim=0 6.2cm 0 0, clip ]{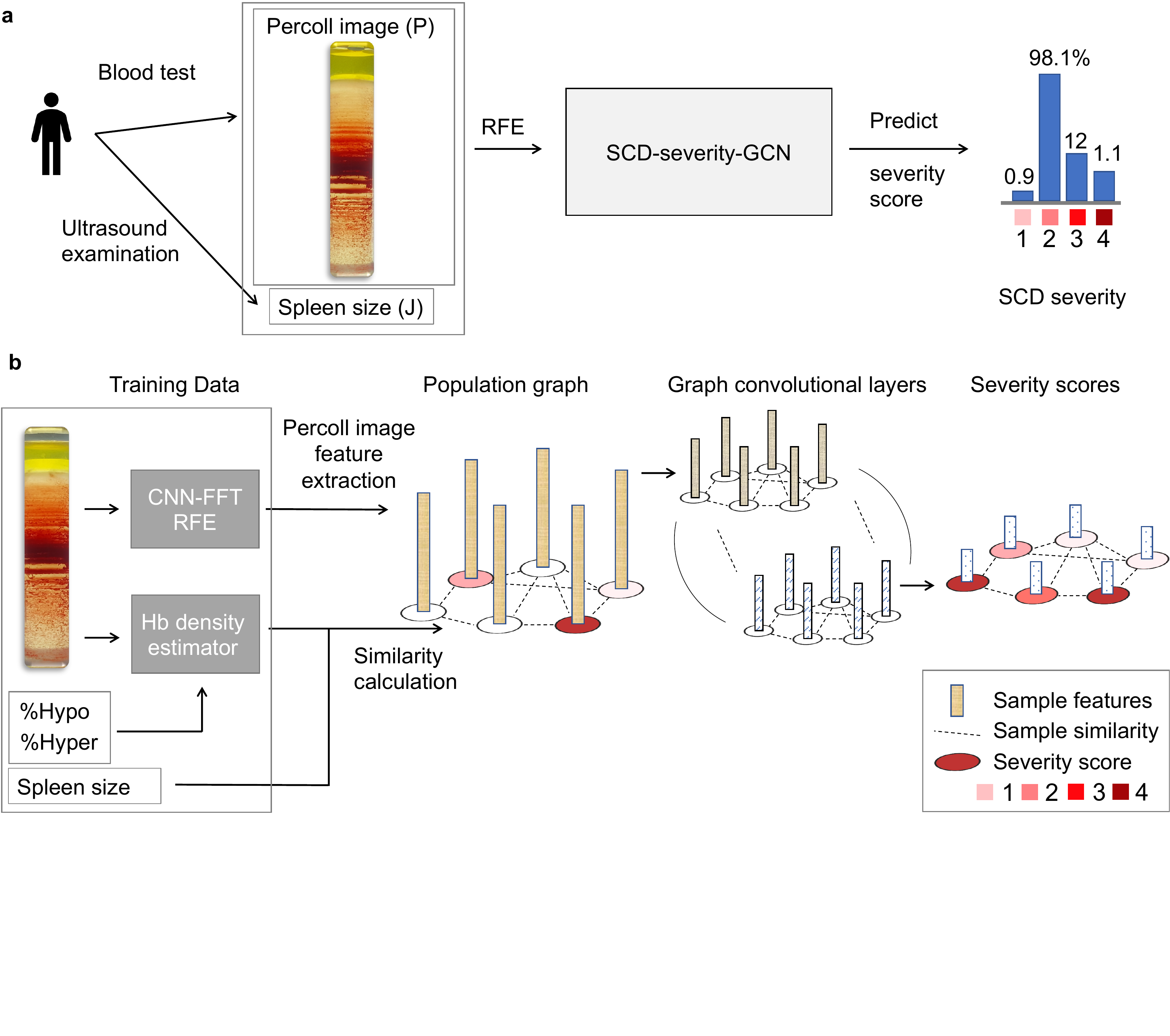}
    \caption{Overview of the proposed SCD-severity-GCN approach. a) A Percoll image from a conventional blood test and the spleen size obtained by ultrasound examination are passed to our trained SCD-severity-GCN to predict a SCD severity score for the patient. 
    b) The SCD-severity-GCN is trained in the following way: Features from the Percoll image are extracted with a convolutional neural network and a fast Fourier transform (CNN-FFT \cite{isbiPercoll}). Another independently trained CNN estimates the hemoglobin (Hb) density of hypochromic (Hypo) and hyperchromic (Hyper) cells. Together with the spleen size of the patient, a similarity measure is calculated between the nodes of a population graph. After two layers of graph convolutions, a severity score for every sample is predicted. 
}
\label{overview}
\end{figure}

Our proposed method, SCD-severity-GCN aims at predicting SCD severity from cheap and easy accessible patient data and consists of the following steps: (i) The abundance of hypo- and hyperchromic cells in the blood sample are predicted based on a Percoll image; (ii) Relevant features are extracted from the Percoll image using a CNN and fast Fourier transform (FFT). (iii) A similarity metric between Percolls based on a patient's spleen size
and the predicted abundance of hypo- and hyperchromic cells is calculated to form a population graph. Using GCNs the SCD severity is predicted (Fig \ref{overview})

\subsection{Model}
Our goal is to have a model $f$ that takes a Percoll image $P_i$ and the spleen size $J_i$ of a patient sample $i$ to return a severity grade $S_i$:
\begin{equation}
    S_i = f(P_i, J_i; \theta)
\end{equation}
where $\theta$ are the model parameters that are learned by training on the dataset.

\subsection{Feature extraction}
For primary feature extraction the approach proposed in \cite{isbiPercoll} is employed.  There, the extraction of Fourier features from the images has been demonstrated to enhance disease classification performance on Percoll images. 
Accordingly, we extract features with an AlexNet \cite{alexnet} architecture and combine them with features from FFT (see Fig. 1). We obtain pretrained weights of the model $f_{\mathrm{cnn-fft}}$ and use the activations preceding the final classification layer as features for our GCN approach.

To reduce feature dimensions, we used recursive feature elimination  \cite{guyon2002gene} and a Ridge classifier as suggested by Parisot et al. \cite{parisot2017spectral}.

\begin{equation}
    x_i = \mathrm{RFE}(f_{\mathrm{cnn-fft}}) (P_i)
\end{equation}
where $x_i$ is the feature vector extracted for the Percoll image $P_i$. 
Also in our approach this step improved the convergence of the training significantly.

\subsection{Graph convolution network}
One of the most intuitive ways of representing populations and their similarities is through graphs.
In our approach, every Percoll image $P_i$ is represented by a vertex $v \in \mathcal{V}$ and the similarity between the Percoll images is modelled by weighted edges $\mathcal{E}$ calculated from the expensive laboratory data (the percentages of hypo- and hyperchromic) which are predicted and cheap clinical data (i.e. spleen size of the patient) (see Fig 2).
A population graph $\mathcal{G} = \{\mathcal{V},\mathcal{E}\}$ is defined accordingly \cite{parisot2017spectral}.

\subsection{Hemoglobin density estimation}
To allow for an application of the method without expensive laboratory testing, the percentages of hypo- and hyperchromic cells $\hat{H}$ in the blood are estimated by a regression. A CNN $f_{chrome}$ is proposed for this task. The groundtruth values $H$ are provided for every Percoll image and are used to train the network:

\begin{equation}
    \mathcal{L}_{chrome} (\gamma) = \frac{1}{N} \sum_{i = 1}^{N} (H_i - \hat{H_i})^2
\end{equation}
where $\hat{H_i} = f_{chrome}(P_i; \gamma)$ and $\gamma$ is the network parameters.

\subsection{Similarity metric}
Under the assumption that patients with similar features experience comparable severity of the disease, the similarity between two samples $v$ and $w$ is calculated via

\begin{equation}
    \mathcal{E}(v_i,v_j) = e^{-(||\hat{H}_{v_i} - \hat{H}_{v_j}|| + \lambda[J_{v_i} == J_{v_j}])}
\end{equation}
where $\hat{H}$ is the vector of estimated percentages of hypo- and hyperchromic cells and $J$ is the spleen size, as above. Iverson brackets yield 1 in case of equality and 0 otherwise. Note that spleen sizes are given as discrete numbers in centimeters (see Fig. 2), obtained in the clinic with a conventional ultrasound device. The coefficient $\lambda$ is set to weight the importance of spleen and lab measurements.

\section{Experiments}

\subsection{Dataset}
\begin{figure}[t]
    \centering
    \includegraphics[width=\textwidth, page=2, trim=0 15cm 0 0, clip ]{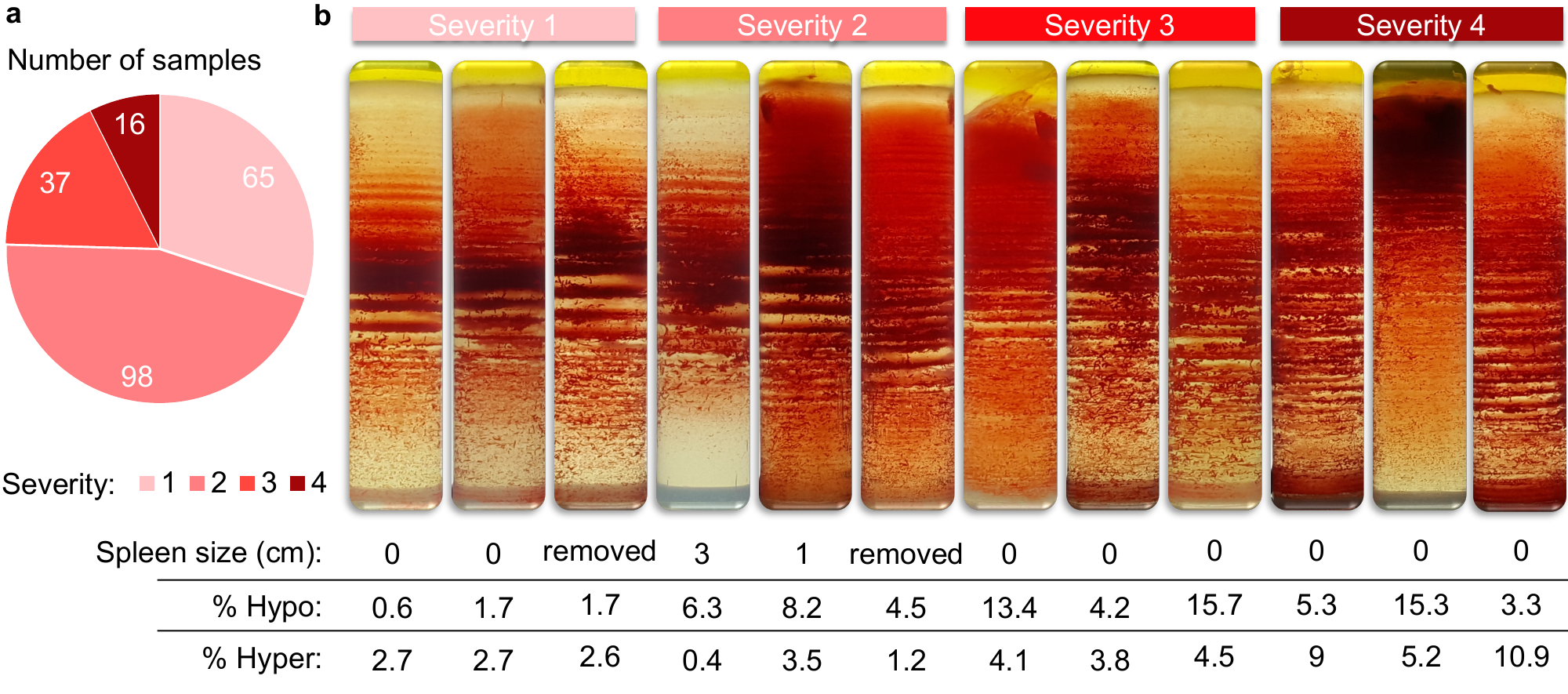}
    \caption{Dataset overview. a) Pie chart shows distribution of severity values for the patients in the dataset. b) Example images sorted according to their severity scores. Corresponding clinical and laboratory test data for each Percoll is also demonstrated. Spleen size of the patients with autosplenectomy and splenectomy is indicated with removed and 0 respectively. }
    \label{fig.dataset}
\end{figure}

Our dataset consists of the 216 samples with Percoll images and laboratory data (\% hypo, \%hyper) and clinical data (spleen size) obtained from 17 patients diagnosed with SCD, who participated in a clinical trial (NCT03247218) conducted in Emek Medical Center in Afula\footnote{https://clinicaltrials.gov/ct2/show/NCT03247218}.
The study has been conducted in accordance with local ethics committee guidelines and the Declaration of Helsinki.
Blood samples were acquired during pre-planned monthly visits according to the trial protocol. For every visit the patient's health was evaluated using blood analysis, including RBC characteristics and measurement of hemolytic and inflammatory markers, urine analysis and blood pressure measurements. Severity of a patient's condition at each measurement point was estimated using the scoring approach proposed by Sebastiani et al. \cite{sebastiani2007network} with minor modifications on disease severity score calculation.
Figure \ref{fig.dataset} shows distribution of severity scores and example samples from the dataset.

\subsection{Implementation details}

\textbf{Hemoglobin density estimation}:
A CNN with seven convolutional layers with ReLU activation function and max-pooling is used. After global average pooling and two fully connected layers the output is regulated with a final ReLU. 
Two dropout layers with a drop rate of $0.5$ are used for regularization.
\\\textbf{Feature extraction}:
The output size from CNN-FFT is 1024, which is reduced with recursive feature elimination (RFE) \cite{guyon2002gene} to 50 features. These features are used as the final feature vector for each Percoll  image.
\\
\textbf{Graph convolutional network}:
A population graph \cite{parisot2017spectral} is created based on the defined feature vectors and similarities. 
We use two hidden layers in the graph and 50 filters in the hidden layers. The dropout rate is set to $0.2$.
For similarity calculation $\lambda$ is set to $10$. 
\\
\textbf{Training}:
Both training procedures are carried out on a 10-fold cross validation dataset.
The model $f_{chrome}$ estimating hemoglobin density is trained with AMSGrad variation of Adam optimizer for $100$ epochs and a learning rate of $0.0005$. 
The graph convolutional network is trained for 300 epochs using Adam optimizer and a learning rate of $0.01$. We use the Tensorflow framework for implementation and training.
\\
\textbf{Evaluation metrics}:
We are reporting root mean square error (RMSE) for the regression task of hemoglobin density estimation. Accuracy, weighted F1-score and area under ROC are reported for the severity grading as well as the area under precision recall curve for every class. Scikit-learn \cite{scikit-learn} implementation is used for calculation all of the metrics.
\\
\textbf{Baseline}:
A linear SVM \cite{scikit-learn} trained on the feature vectors is used as a baseline for our grading approach.

\subsection{Results}
The dataset is divided into 10 stratified folds for patient-wise cross validation. All of the models are independently ran on each combination of these folds. Mean and standard deviation is reported for all of the 10 experiments.

First, the values predicted by the Hb density estimation model $f_{chrome}$ based on Percoll images are compared against the actual lab tests. The root mean square error (RMSE) of the percentage for hypochromic cells is $6.5 \pm 4.0$ and for hyperchromic cells $0.90 \pm 0.12$. Considering the ranges of the hypo and hyper values, which are  $[0.6,  37.5]$ and $[0.2, 10.9]$, respectively, we consider the estimation sufficiently good. 

\begin{figure}
    \centering
    \includegraphics[width=.7\textwidth, page=3, trim=0 18.7cm 10cm 0, clip ]{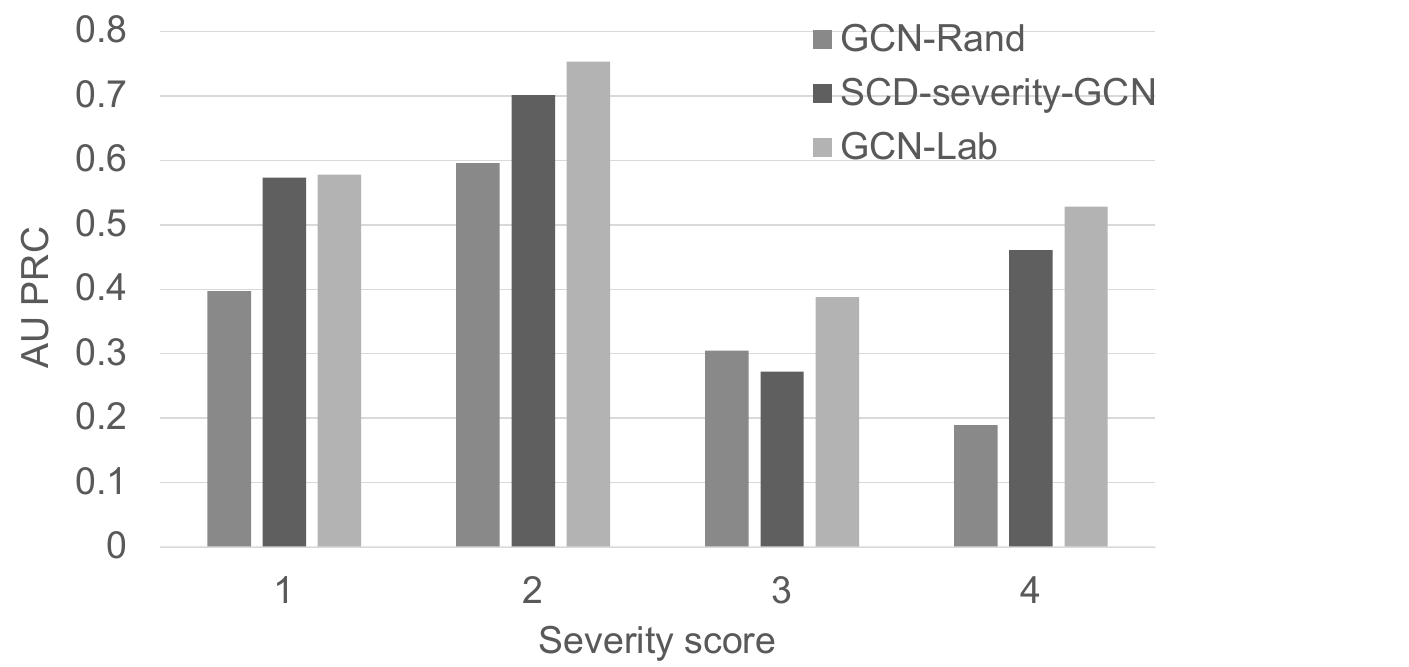}
    \caption{Area under precision recall curve for different methods per class.}
    \label{fig_auprc}
\end{figure}

Next, we compare our SCD-severity-GCN approach with the following methods: (i) A linear SVM trained on the $x_i$ features vectors extracted from the Percoll image (SVM), (ii) a linear SVM trained on $x_i$ feature vectors and the cheap clinical ultrasound and newly proposed and time consuming groundtruth lab information (SVM - Lab), (iii) a GCN based on randomized laboratory information (GCN - Rand), and (iv) a GCN using not the estimated, but the actual laboratory information (GCN - Lab) as the upper limit.
Table \ref{tab.results} shows that our SCD-severity-GCN approach using estimated Hb densities is close to the GCN that required hard to obtain lab data (GCN - Lab) in terms of accuracy, weighted F1-score and area under ROC.
Since the dataset is unbalanced, we are reporting the area under precision recall curve in Figure \ref{fig_auprc} for every class and different approaches.  
\begin{table}[t]
\centering
\caption{Our proposed SCD-severity-GCN    method based on GCN outperforms other approaches. Prediction of hypo- and hyperchromic cells percentage slightly affects the performance while providing independence from expensive cell counters. }
\label{tab.results}
\begin{tabular}{p{3cm}|P{2.5cm}|P{2.5cm}|P{2.5cm}}

 & Accuracy & F1 - Score & AU ROC \\ \hline
SVM & 0.44 ± 0.07 & 0.28 ± 0.02 & - \\ \hline
% SVM + & 0.46 ± 0.07 & 0.34 ± 0.04 & - \\ \hline
SVM - Lab. & 0.39 ± 0.14 & 0.29 ± 0.08 & - \\ \hline
GCN - Rand. & 0.53 ± 0.05 & 0.42 ± 0.08 & 0.53 ± 0.20 \\ \hline
SCD-severity-GCN & 0.61 ± 0.13 & 0.53 ± 0.17 & 0.61 ± 0.25 \\ \hline
GCN - Lab. & 0.65 ± 0.15 & 0.59 ± 0.19 & 0.67 ± 0.24
\end{tabular}

\end{table}

\subsection{Ablation study}
GCNs are sensitive to the formulation of the graph adjacency matrix based on the pairwise similarity that is defined between the nodes.
Choosing parameters that are biologically significant and easy to obtain is crucial.  
To evaluate the importance of the different clinical (spleen size) and laboratory (\% of hypo- and hyperchromic cells) information used for the formation of our GCN, we designed an ablation study and compare GCNs trained with different combinations of these parameters.
As Table 2 shows, the combination based on spleen size and percentages of hypo- and hyperchromic RBCs yields the best result.

\begin{table}
\caption{Combination of different clinical and laboratory measurements results in a slightly different shape for the GCN and thus different performance.}
\begin{tabular}{p{4.5cm}|P{2.4cm}|P{2.4cm}|P{2.4cm}}
 GCN similarity parameters& Accuracy & F1 - Score & AU ROC \\ \hline
Spleen & 0.45 ± 0.01 & 0.28 ± 0.01 & 0.48 ± 0.18 \\ \hline
Spleen \& Hypo & 0.63 ± 0.11 & 0.56 ± 0.14 & 0.63 ± 0.23 \\ \hline
Spleen \& Hyper & 0.62 ± 0.16 & 0.54 ± 0.21 & 0.60 ± 0.25 \\ \hline
Hypo \& Hyper & 0.55 ± 0.08 & 0.46 ± 0.12 & 0.55 ± 0.22 \\ \hline
Spleen \& Hypo \& Hyper & \textbf{0.65 ± 0.15} & \textbf{0.59 ± 0.19} & \textbf{0.67 ± 0.24}
\end{tabular}
\end{table}

\subsection{Discussion}
\begin{figure}[t]
    \centering
    \includegraphics[width=\textwidth, page=4, trim=0 20cm 0 0, clip ]{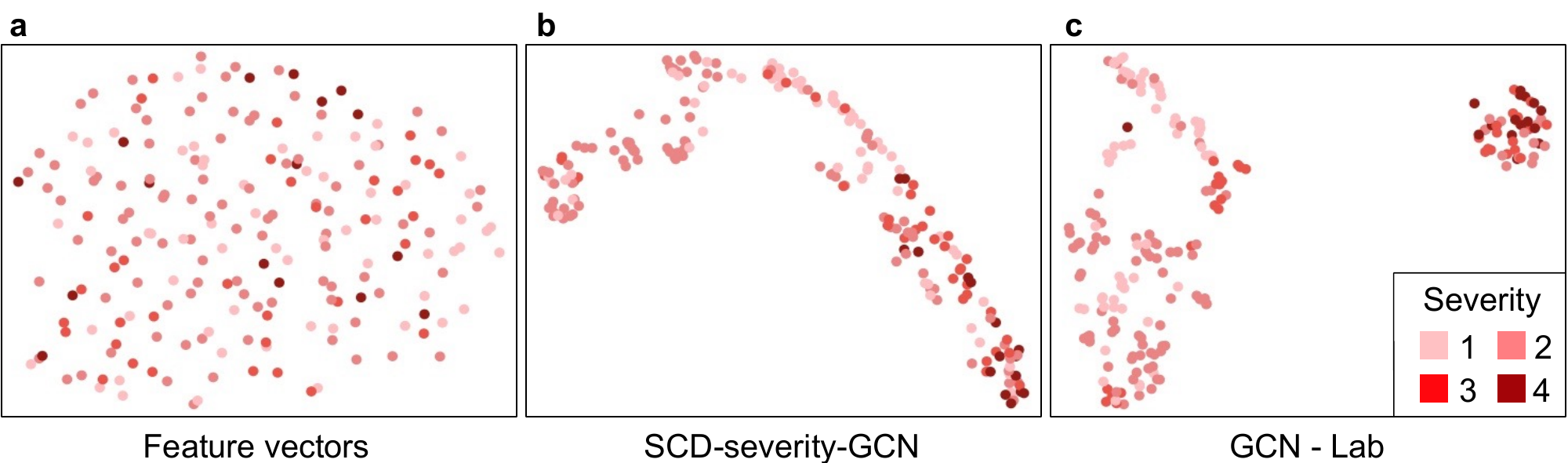}
    \caption{UMAP embedding of the (a) processed feature vectors, (b) the GCN with predicted ones and (c) the GCN with groundtruth lab information. Clear separation of samples with different severity scores by the proposed GCN is evident.}
    \label{fig.umap}
\end{figure}

Severity prediction of SCD is a challenging task normally preformed with several clinical and laboratory tests. Here we propose a novel potential severity prediction approach based on RBC density separation (as provided by Percoll gradients) analysis that may amend the currently existing ones. Information obtained solely from Percoll images is not be sufficient for an acceptable classification  (see Table \ref{tab.results}), even though those features sufficed for successful diagnosis of different anemias \cite{isbiPercoll}. By combining Percoll derived features with complementary clinical and laboratory data and training a GCN with this information, we can achieve an accuracy that is surprisingly high for this challenging clinical task. 
This is illustrated by the UMAP embedding of feature vectors (Fig \ref{fig.umap} a), and GCN outputs with estimated (Fig \ref{fig.umap} b) and groundtruth lab information (Fig \ref{fig.umap} c). Samples from different severity classes are nicely disentangled in the UMAP thanks to the GCN approach we utilized. Although clustering using the groundtruth lab information (GCN - Lab) is a lot better, a smooth transition from low to high severity is already evident in the approach that uses estimated Hb density only (SCD-severity-GCN).

\section{Conclusion}
Sickle cell disease severity prediction is an important task that allows to prevent life-threatening complications, reduce morbidity and mortality and refine the choice of optimal therapeutic strategies \cite{sebastiani2007network}. Offering affordable and versatile solutions for improving life quality of the SCD patients is a necessity specially in low resource areas of the planet. 
Here, we proposed the first computational method requiring only the Percoll gradient image and spleen size obtained from a conventional ultrasound. Analysis of Percoll gradient images with CNNs nicely predicted percentages of hypo- and hyperchromic cells and the proposed GCN predicted SCD severity score with a surprisingly high accuracy. Our approach uses a unique combination of methods, with a GCN at its heart.

Results look very promising and provide a solid ground for future work. Next we will analyze more patients, especially more severe ones as well as pediatric datasets. Our SCD-severity-GCN based on Percoll  images  requires much smaller volumes of blood  compared to common hematological tests (1 ml or less instead of 7-10 ml), which is particularly relevant for kids and patients suffering from severe anemia.

\section*{Acknowledgments}
Special thanks to Prof. Ariel Koren and Dr. Carina Levin from the Emek Medical Center in Afula who made this work possible. This project has received funding from the European Union’s Horizon 2020 research and innovation programme under grant agreement No 675115 — RELEVANCE — H2020-MSCA-ITN-2015/ H2020-MSCA-ITN-2015. The work of L.L. was funded by UZH Foundation.
C.M. and A.S. have received funding from the European Research Council (ERC) under the European Union’s Horizon 2020 research and innovation programme (Grant agreement No. 866411). 

\bibliographystyle{splncs04}
\bibliography{article}
\end{document}